\documentclass[proceedings]{JHEP3}

\PrHEP{PrHEP hep2001}			
\conference{International Europhysics Conference on HEP}		

\title{CP Violation in a SUSY $SO(10) \times U(2)_{F}$ Model
\thanks{Talk presented by K.T. Mahanthappa at EPS
  HEP2001, Budapest, Hungary, 12 - 18 July, 2001, and by M.-C. Chen at Snowmass
  2001, Snowmass Village, CO, 30 June - 21 July, 2001.}}

\author{Mu-Chun Chen and %
        \speaker{K.T. Mahanthappa}\\%
        Department of Physics, University of Colorado, Boulder, %
        CO 80309-0390, U.S.A.\\%
        E-mail: \email{mu-chun.chen@colorado.edu}, %
                \email{ktm@verb.colorado.edu} }  

\abstract{We construct a model based on SUSY $SO(10)$ combined with $U(2)$
family symmetry including complex phases leading to CP violation. 
In contrast with the commonly used effective operator approach,
$\overline{126}$-dimensional Higgs fields are utilized to construct the Yukawa
sector. $R$-parity symmetry is thus preserved at low energies. The {\it
symmetric} mass textures arising from the left-right symmetry breaking chain
of $SO(10)$ give rise to very good predictions for quark and lepton masses and
mixings. The prediction for $\sin 2\beta$ agrees with the current bounds from
BaBar and Belle. In the neutrino sector, our predictions are in good agreement
with results from atmospheric neutrino experiments. Our model favors both the
LOW and QVO solutions to the solar neutrino anomaly; the matrix element for
neutrinoless double beta decay is highly suppressed. The leptonic analog of
the Jarlskog invariant, $J_{CP}^{l}$, is predicted to be of $O(10^{-2})$.}

\begin{document}
$SO(10)$ has long been thought to be an attractive candidate for a
grand unified theory (GUT) for a number of reasons: First of all, it unifies
all the $15$ known fermions with the right-handed neutrino for each family
into one $16$-dimensional spinor representation. The seesaw mechanism then
arises very naturally, and the non-zero neutrino masses can thus be explained.
Since a complete quark-lepton symmetry is achieved, it has the promise for
explaining the pattern of fermion masses and mixing. Because $B-L$ contained
in $SO(10)$ is broken in symmetry breaking chain to the SM, it also has the
promise for baryogenesis. Recent atmospheric neutrino oscillation data from
Super-Kamiokande indicates non-zero neutrino masses. This in turn gives very
strong support to the viability of $SO(10)$ as a GUT group. Models based on
$SO(10)$ combined with discrete or continuous family symmetry have been
constructed to understand the flavor problem. Most of the models utilize 
``lopsided'' mass textures which usually require
more parameters and therefore are less constrained. Furthermore, the
right-handed neutrino Majorana mass operators in most of these models are made
out of $16_{H} \times 16_{H}$ which breaks the $R$-parity at a very
high scale. The aim of this talk, based on Ref.\cite{cm,cm2}, 
is to present a realistic model based on supersymmetric $SO(10)$ combined
with $U(2)$ family symmetry which successfully predicts the low energy
fermion masses and mixings. Since we utilize {\it symmetric} mass textures and
$\overline{126}$-dimensional Higgs representations for the right-handed
neutrino Majorana mass operator, our model is more constrained in addition to
having $R$-parity conserved. We first discuss the viable phenomenology of mass
textures followed by the model which accounts for it, and then the implications
of the model for neutrino mixing, CP violation, neutrinoless double beta decay 
and leptogenesis are presented.

The set of up- and down-quark mass matrix combination is given by,
at the GUT scale,
\begin{equation}
\label{eq:Mud}
M_{u}=\left(
\begin{array}{ccc}
0 & 0 & a e^{i\gamma_{a}} \\
0 & b e^{i\gamma_{b}} & c e^{i\gamma_{c}} \\
a e^{i\gamma_{a}} & c e^{i\gamma_{c}} & e^{i\gamma}
\end{array}
\right) d v_{u}, \qquad 
M_{d}=\left(
\begin{array}{ccc}
0 & e e^{i\gamma_{e}} & 0 \\
e e^{i\gamma_{e}} & f e^{i\gamma_{f}} & 0 \\
0 & 0 & e^{i\gamma_h}
\end{array}
\right) h v_{d}
\end{equation}
with $a \simeq b \ll c \ll 1$, and $e \ll f \ll 1$.
Symmetric mass textures arise naturally if $SO(10)$ breaks down to the SM
through the left-right symmetric breaking chain $SU(4) \times SU(2)_{L}
\times SU(2)_{R}$. $SO(10)$ relates the up-quark mass matrix to 
the Dirac neutrino mass matrix, and the down-quark mass matrix to 
the charged lepton mass matrix. To achieve the Georgi-Jarlskog relations, 
$m_{d} \simeq 3 m_{e}$, $m_{s} \simeq \frac{1}{3} m_{\mu}$, 
$m_{b} \simeq m_{\tau}$, 
a factor of $-3$ is needed
in the $(2,2)$ entry of the charged lepton mass matrix,  
\begin{equation}
\label{eq:Me}
M_{e} = \left(
\begin{array}{ccc}
0 & e e^{i\gamma_{e}} & 0 \\
e e^{i\gamma_{e}} & -3 f e^{i\gamma_{f}} & 0 \\
0 & 0 & e^{i\gamma_h}
\end{array}
\right) h v_{d}
\end{equation}
This factor of $-3$ can be accounted for by the $SO(10)$ CG 
coefficients associated with $\overline{126}$-dimensional
Higgs representations.  
In order to explain the smallness of the neutrino masses, we will adopt
the type I seesaw mechanism. 
The Dirac neutrino mass matrix is identical to
the mass matrix of the up-quarks in the framework of $SO(10)$   
\begin{equation}
\label{eq:Mlr}
M_{\nu_{LR}} = \left(
\begin{array}{ccc}
0 & 0 & a e^{i\gamma_{a}} \\
0 & b e^{i\gamma_{b}} & c e^{i\gamma_{c}} \\
a e^{i\gamma_{a}} & c e^{i\gamma_{c}} & e^{i\gamma}
\end{array}
\right) d v_{u}
\end{equation}
The right-handed neutrino sector is an unknown sector. It is only constrained
by the requirement that it gives rise to a bi-maximal mixing pattern and a
hierarchical mass spectrum at low energies. To achieve this, we consider an
effective neutrino mass matrix of the form
\begin{equation}
\label{eq:Mll}
M_{\nu_{LL}}=M_{\nu_{LR}}^{T} M_{\nu_{RR}}^{-1} M_{\nu_{LR}}
= \left( 
\begin{array}{ccc}  
0 & 0 & t \\
0 & 1 & 1 \\  
t & 1 & 1 
\end{array} \right) \frac{d^{2}v_{u}^{2}}{M_{R}} 
\end{equation}
The effective neutrino mass matrix of this form is obtained 
if the right-handed neutrino mass matrix has the same texture as that of the
Dirac  neutrino mass matrix, 
\begin{equation} 
\label{eq:Mrr}
M_{\nu_{RR}}=
\left( \begin{array}{ccc}
0 & 0 & \delta_{1} \\
0 & \delta_{2} & \delta_{3} \\
\delta_{1} & \delta_{3} & 1
\end{array} \right) M_{R}
\end{equation}
and if the elements $\delta_{i}$ are of the right orders of magnitudes, 
determined by $\delta_{i}=f_{i}(a,b,c,t,\theta)$, where 
$\theta \equiv (\gamma_{b}-2\gamma_{c}-\gamma)$.
Note that $M_{\nu_{LL}}$ has the same texture as that of $M_{\nu_{LR}}$ and
$M_{\nu_{RR}}$. That is to say, the seesaw mechanism is form invariant.  A
generic feature of mass matrices of the type given in Eq.(\ref{eq:Mll}) 
is that they give rise to 
bi-maximal mixing pattern. After diagonalizing this mass matrix, one can see
immediately that the squared mass difference between $m_{\nu_{1}}^{2}$ and
$m_{\nu_{2}}^{2}$ is of the order of
$O(t^{3})$, while the squared mass difference between 
$m_{\nu_{2}}^{2}$ and $m_{\nu_{3}}^{2}$ 
is of the order of $O(1)$, in units of $\Lambda$. 
For $t \ll 1$, the phenomenologically favored relation $\Delta
m_{atm}^{2} \gg \Delta m_{\odot}^{2}$ is thus obtained.

The $U(2)$ family symmetry is implemented {\it {\'a} la} the
Froggatt-Nielsen mechanism 
which simply states that the heaviest matter fields acquire their masses through
tree level interactions with the Higgs fields while masses of the lighter
matter fields are produced by higher dimensional interactions involving, in
addition to the regular Higgs fields, exotic vector-like pairs of matter
fields and the so-called flavons (flavor Higgs fields). After integrating out
superheavy  $(\approx M)$ vector-like matter fields, the mass terms of the
light matter fields get suppressed by a factor of $\frac{<\theta>}{M}$, where
$<\theta>$ is the VEVs of the flavons and $M$ is the UV-cutoff of the
effective theory above which the family symmetry is exact. 
We assume that the family symmetry scale is higher than the GUT scale. 
The heaviness of the top quark and to suppress the
SUSY FCNC together suggest that the third family of matter fields transform
as a singlet and the lighter two families of matter fields transform as a
doublet under $U(2)$. In the family symmetric limit, only the third family has
non-vanishing Yukawa couplings. $U(2)$ breaks down in two steps:
$U(2) \stackrel{\epsilon M}{\longrightarrow} 
U(1) \stackrel{\epsilon' M}{\longrightarrow}
nothing$, 
where $\epsilon' \ll \epsilon \ll 1$ and $M$ is the family symmetry scale. 
These small parameters $\epsilon$ and $\epsilon'$ are the ratios of 
the vacuum expectation
values of the flavon fields to the family symmetry scale. 
Since $\psi_{3}\psi_{3} \sim 1_{S}$, 
$\psi_{3}\psi_{a} \sim 2$, $\psi_{a}\psi_{b} \sim 2 \otimes 2 = 1_{A} \oplus 3$, 
the only relevant flavon fields are in the 
$A^{ab} \sim 1_{A}$, $\phi^{a} \sim 2$, and 
$S^{ab} \sim 3$ dimensional representations of $U(2)$. 
Because we are confining ourselves to symmetric mass textures, we use only
$\phi^{a}$ and $S^{ab}$. 
In the chosen basis, the VEVs various flavon fields could acquire are given
by 
\begin{equation}
\frac{ \left< \phi \right> }{M} \sim O \left( \begin{array}{c}
\epsilon' \\
\epsilon
\end{array} \right), \quad
\frac{ \left< S^{ab} \right> }{M} \sim O \left( \begin{array}{cc}
\epsilon' & \epsilon' \\
\epsilon' & \epsilon
\end{array} \right)
\end{equation}
Putting everything together, a symmetric mass matrix would 
have the following built-in
hierarchy given by
\begin{equation}
\left( \begin{array}{ccc}
\epsilon' & \epsilon' & \epsilon' \\
\epsilon' & \epsilon & \epsilon \\
\epsilon' & \epsilon & 1 
\end{array} \right)
\end{equation}
Combining $SO(10)$ with $U(2)$, the most general
superpotential which respects the symmetry 
one could write down is given schematically by
\begin{equation}
W = H(\psi_{3}
\psi_{3} + \psi_{3} \frac{\phi^{a}}{M} \psi_{a} + \psi_{a}\frac{S^{ab}}{M}
\psi_{b}) 
\end{equation}

A discrete symmetry $(Z_{2})^{3}$ is needed to avoid unwanted couplings.
The field content of our model is then given by\\
-- matter fields
\begin{displaymath}
\psi_{a} \sim (16,2)^{-++} \quad (a=1,2), \qquad
\psi_{3} \sim (16,1)^{+++}
\end{displaymath}
-- Higgs fields:
\begin{displaymath}
\begin{array}{ll}
(10,1):& \quad T_{1}^{+++}, \quad T_{2}^{-+-},\quad
T_{3}^{--+}, \quad T_{4}^{---}, \quad T_{5}^{+--}\\
(\overline{126},1): & \quad \overline{C}^{---}, \quad \overline{C}_{1}^{+++},
\quad \overline{C}_{2}^{++-}
\end{array}
\end{displaymath}
-- Flavon fields:
\begin{displaymath}
\begin{array}{ll}
(1,2): & \quad \phi_{(1)}^{++-}, \quad \phi_{(2)}^{+-+}, \quad \Phi^{-+-}
\\
(1,3): & \quad S_{(1)}^{+--}, \quad S_{(2)}^{---}, \quad
\Sigma^{++-}
\end{array}
\end{displaymath}
and the superpotential of our model which generates fermion masses is given by
\begin{equation}
\begin{array}{l}
W = W_{D(irac)} + W_{M(ajorana)}
\\
W_{D}=\psi_{3}\psi_{3} T_{1} + \frac{1}{M} \psi_{3} \psi_{a}
\left(T_{2}\phi_{(1)}
+T_{3}\phi_{(2)}\right)
+ \frac{1}{M} \psi_{a} \psi_{b} \left(T_{4} + \overline{C}\right) S_{(2)}
+ \frac{1}{M} \psi_{a} \psi_{b} T_{5} S_{(1)}
\\
W_{M}=\psi_{3} \psi_{3} \overline{C}_{1} 
+ \frac{1}{M} \psi_{3} \psi_{a} \Phi \overline{C}_{2}
+ \frac{1}{M} \psi_{a} \psi_{b} \Sigma \overline{C}_{2}
\end{array}
\end{equation}
where $T_{i}$'s and $\overline{C}_{i}$'s are the $10$ and $\overline{126}$
dimensional Higgs representations of $SO(10)$ respectively, and $\Phi$ and 
$\Sigma$ are the 
doublet and triplet of $U(2)$, respectively. 
Detailed quantum number assignment
and the VEVs acquired by various scalar fields are given in Ref.\cite{cm}. This
superpotential gives rise to the mass textures given in 
Eq.(\ref{eq:Mud})-(\ref{eq:Mrr}):
\begin{equation}
M_{u,\nu_{LR}}=
\left( \begin{array}{ccc}
0 & 0 & \left<10_{2}^{+} \right> \epsilon'\\
0 & \left<10_{4}^{+} \right> \epsilon & \left<10_{3}^{+} \right> \epsilon \\
\left<10_{2}^{+} \right> \epsilon' & \left<10_{3}^{+} \right> \epsilon &
\left<10_{1}^{+} \right>
\end{array} \right)
= 
\left( \begin{array}{ccc}
0 & 0 & r_{2} \epsilon'\\
0 & r_{4} \epsilon & \epsilon \\
r_{2} \epsilon' & \epsilon & 1
\end{array} \right) M_{U}
\end{equation}
\begin{equation}
M_{d,e}=
\left(\begin{array}{ccc}
0 & \left<10_{5}^{-} \right> \epsilon' & 0 \\
\left<10_{5}^{-} \right> \epsilon' &  (1,-3)\left<\overline{126}^{-} \right>
\epsilon & 0\\ 0 & 0 & \left<10_{1}^{-} \right>
\end{array} \right)
=
\left(\begin{array}{ccc}
0 & \epsilon' & 0 \\
\epsilon' &  (1,-3) p \epsilon & 0\\
0 & 0 & 1
\end{array} \right) M_{D}
\end{equation}
where
$M_{U} \equiv \left<10_{1}^{+} \right>$, 
$M_{D} \equiv \left<10_{1}^{-} \right>$, 
$r_{2} \equiv \left<10_{2}^{+} \right> / \left<10_{1}^{+} \right>$, 
$r_{4} \equiv \left<10_{4}^{+} \right> / \left<10_{1}^{+} \right>$ and
$p \equiv \left<\overline{126}^{-}\right> / \left<10_{1}^{-} \right>$.
The right-handed neutrino mass matrix is  
\begin{equation}
M_{\nu_{RR}}=  
\left( \begin{array}{ccc}
0 & 0 & \left<\overline{126}_{2}^{'0} \right> \delta_{1}\\
0 & \left<\overline{126}_{2}^{'0} \right> \delta_{2} 
& \left<\overline{126}_{2}^{'0} \right> \delta_{3} \\ 
\left<\overline{126}_{2}^{'0} \right> \delta_{1}
& \left<\overline{126}_{2}^{'0} \right> \delta_{3} &
\left<\overline{126}_{1}^{'0} \right> \end{array} \right)
= 
\left( \begin{array}{ccc}
0 & 0 & \delta_{1}\\
0 & \delta_{2} & \delta_{3} \\ 
\delta_{1} & \delta_{3} & 1
\end{array} \right) M_{R}
\label{Mrr}
\end{equation}
with $M_{R} \equiv \left<\overline{126}^{'0}_{1}\right>$.
Note that, since we use
$\overline{126}$-dimensional representations of Higgses to generate the heavy
Majorana neutrino mass terms, R-parity is preserved at all energies. 

With values of $m_{f}, (f= u,c,t,e,\mu,\tau)$ and those of $|V_{us, ub, cb}|$
at the weak scale, the input parameters at the GUT scale are determined. The
predictions for the charged fermion masses and CKM mixing 
of our model at $M_{Z}$ which are summarized
in Table[\ref{table1}] including 2-loop RGE effects are in good agreements
with the experimental values. 
\TABLE{{\tiny
\begin{tabular}
{l c | c c l c c c l}
\hline
 & & & & experimental results \qquad \qquad
 & & & & predictions at $M_{z}$ \\
& & & & extrapolated to $M_{Z}$ \qquad \qquad
 & & & & \\
\hline
$\frac{m_{s}}{m_{d}}$
& & & & $17 \sim 25$
& & & & $25$\\
$m_{s}$
& & & & $93.4^{+11.8}_{-13.0}MeV$
& & & & $85.66 MeV$\\
$m_{b}$
& & & & $3.00^{+}_{-}0.11GeV$
& & & & $3.147 GeV$\\
\hline
$\vert V_{ud} \vert$
& & & & $0.9745-0.9757$
& & & & $0.9751$\\
$\vert V_{cd} \vert$
& & & & $0.218-0.224$
& & & & $0.2218$\\
$\vert V_{cs} \vert$
& & & & $0.9736-0.9750$
& & & & $0.9744$\\
$\vert V_{td} \vert$
& & & & $0.004-0.014$
& & & & $0.005358$\\
$\vert V_{ts} \vert$
& & & & $0.034-0.046$
& & & & $0.03611$\\
$\vert V_{tb} \vert$
& & & & $0.9989-0.9993$
& & & & $0.9993$ \\
$J_{CP}^{q}$
& & & & $(2.71^{+}_{-}1.12) \times 10^{-5}$
& & & & $1.748 \times 10^{-5}$ \\
$\sin 2\alpha$
& & & & $-0.95 \; - \; 0.33$
& & & & -0.8913\\
$\sin 2\beta$
& & & & $0.59^{+}_{-}0.14^{+}_{-}0.05$ (BaBar)
& & & & $0.7416$ \\
& & & & $0.58^{+0.32+0.09}_{-0.34-0.10}$(Belle)
& & & &\\
$\gamma$
& & & & $34^{0}-82^{0}$
& & & & $34.55^{0} \; (0.6030 rad)$\\
\hline
\end{tabular}%
\caption{The predictions for the charged fermion masses, the
CKM matrix elements and the CP violation measures.}\label{table1}}}
In the neutrino sector, the LOW solution to the solar neutrino problem is
obtained with 
$(\delta_{1}, \delta_{2}, \delta_{3}, M_{R}) =
(0.00125,2.22\times 10^{-4}e^{i(0.22)},
0.0165e^{-i(0.0017)},2.01\times10^{13}GeV)$. 
The atmospheric and solar squared
mass differences are predicted to be  $\Delta m_{23}^{2} = 2.95 \times 10^{-3}
eV^{2}$ and  $\Delta m_{12}^{2} = 1.77 \times 10^{-7} eV^{2}$; the mixing
angles are given by 
$\sin^{2} 2\theta_{atm}=0.999$, and 
$\sin^{2} 2\theta_{\odot}= 0.994$. 
$|U_{e \nu_{3}}|$ is predicted to be $0.0750$ which is below the
upper bound $0.16$ by the CHOOZ experiment.
The leptonic Jarlskog invariant is predicted to be $J_{CP}^{l}=-0.00815$, and the
matrix element for the neutrinoless double beta decay is predicted to be
$|<m>|=1.36\times 10^{-3}eV$. The masses of the three heavy neutrinos are
given by $(M_{1},M_{2},M_{3})=(9.41\times 10^{7},1.49\times 10^{9}, 2.01\times
10^{13})GeV$.  We can also have the QVO solution with
$(\delta_{1},\delta_{2},\delta_{3},M_{R})=(0.00127,3.64\times10^{-5}e^{i(0.220)},
0.0150e^{-i(0.0107)},1.22\times 10^{14}GeV)$. In this case, 
$\Delta m_{23}^{2} = 3.12 \times 10^{-3} eV^{2}$, and 
$\Delta m_{12}^{2} = 7.58 \times 10^{-10} eV^{2}$. The mixing angles are given
by 
$\sin^{2} 2\theta_{atm}=0.999$, and 
$\sin^{2} 2\theta_{\odot}=0.995$. 
$|U_{e \nu_{3}}|$ is predicted to be $0.0531$.
$J_{CP}^{l}$ and $<m>$ are predicted to be $-0.00811$ and $3.07\times 10^{-4}eV$
respectively. The masses of the three heavy neutrinos are given by
$(M_{1},M_{2},M_{3})=(3.70\times 10^{7},2.34\times 10^{9}, 1.22\times 10^{14})GeV$.

A few words concerning baryonic asymmetry are in order. Even though the
sphaleron effects destroy baryonic asymmetry, it could be produced as an
asymmetry in the generation of $(B-L)$ at a high scale because of lepton
number violation due to the decay of heavy right-handed Majorana
neutrinos, which in turn is converted into baryonic asymmetry due to
sphalerons. But in our model this mechanism produces baryonic asymmetry of
$O(10^{-13})$ which is too small to account for the observed value of
$(1.7-8.3) \times 10^{-11}$, reasons being that the mass of the lightest
right-handed Majorana neutrino is too small {\it and} the $1-3$ family mixing
of right-handed neutrinos is too large, leading, in essence, to the violation
of the out-of-equilibrium condition required by Sakharov. So a mechanism other
than leptogenesis is required to explain baryonic asymmetry.

This work was supported, in part, by the U.S. Department of Energy under
Grant No. DE FG03-05ER40892.



\end{document}